\documentclass[12pt]{iopart}
 \usepackage{graphicx}
 \usepackage{color}

\begin{document}


\title {Near-Zero Modes in Superconducting Graphene}
\author{Pouyan Ghaemi$^{1,2}$ and Frank Wilczek${^3}$}
\address{$^1$ Department of Physics, University of California at Berkeley, Berkeley, CA 94720}
\address{$^2$ Materials Sciences Division, Lawrence Berkeley National Laboratory, Berkeley, CA 94720}

\address{$^3$ Center for Theoretical Physics, Department of Physics, Massachusetts Institute of Technology, Cambridge, MA 02139}



\begin{abstract}
Vortices in the simplest superconducting state of graphene contain very low energy excitations, whose existence is connected to an index theorem that applies strictly to an approximate form of the relevant Bogoliubov-deGennes equations.   When Zeeman interactions are taken into account, the zero modes required by the index theorem are (slightly) displaced.  Thus the vortices acquire internal structure, that plausibly supports interesting dynamical phenomena.
\end{abstract}


\maketitle

\section{Introduction}
In this paper we will draw together several lines of thought.
Electronic properties of the two-dimensional material graphene have
attracted theoretical interest for many years \cite{oldGrapheneLit},
and of course recently \cite{grapheneReviews}.   Many of graphene's
unusual properties derive from the fact that its conduction and
valence bands touch at two points, forming conical energy surfaces
near those points.  At neutral filling, the Fermi energy coincides
with the apex of the cone.  As we shall discuss momentarily, one can
induce superconductivity in graphene, probably in several forms
\cite{grapheneSuper}.  Even the simplest such superconducting state
has been found to have unusual properties with respect to Andreev
reflection \cite{beenakker}.   Here we point out that vortices, or
more generally multivortices, in this state acquire interesting
internal structure.   This occurs because each vortex supports a
low-energy mode of the equation for electronic excitations, i.e. the
Bogoliubov-deGennes (BdG) equations \cite{deGennes}.  Indeed, an
approximate form of the BdG equation maps, after appropriate
identifications, to an equation of relativistic field theory that
has been investigated by Jackiw and Rossi \cite{jr}, who discovered
zero-energy solutions.  E. Weinberg \cite{eWeinberg} subsequently
demonstrated that the existence of these zero-energy solutions is
connected to an index theorem.    

The structured vortices resemble
in some respects vortices in $p+ip$ superconductors \cite{pip} or in
the Pfaffian quantum Hall state \cite{pfaffian}.  On the surface of topological insulators \cite{moore} there is a single Dirac band, and the index theory discussed here predicts the presence of single Majorana mode in the induced vortex core, as was studied previously by other means \cite{fukane}.  That single Majorana mode imparts a form of nonabelian statistics to the vortices.  In graphene, however, the presence of valley and spin quantum numbers, each two valued, leads to appearance of even number of in-gap modes in the vortex core.   The doubled modes of graphene do not lead to such exotic quantum statistics, but plausibly \cite{ghaemiryulee} they will support the phenomenon of deconfined quantum criticality \cite{dqc}, which has not yet been observed experimentally.  




\section{Induced Superconducivity in Graphene}


At neutral filling the density of states at the Fermi surface, which
degenerates to two points, is very small, and in two dimensions
fluctuations are important \cite{mermin-wagner}, so the prospect for
{\it intrinsic\/} superconductivity in undoped graphene is problematic.
If, however a graphene layer is put in contact with a superconducting
substrate, then electron-electron interactions can induce anomalous
(non-conserving) terms in the effective Hamiltonian, according to
the general scheme
\begin{eqnarray} \label{interaction}
{\cal H}= & - g \Psi^\dagger \Psi^\dagger \Psi \Psi + \ {\rm h. c}. \nonumber \\ \rightarrow  &  -g \langle \Psi^\dagger \Psi^\dagger \rangle  \psi \psi +  {\rm h. c}. \equiv -g \Delta_0^\dagger  \psi \psi +  {\rm h. c}.
\end{eqnarray}
where $\Psi$ is the total electron field, $\Delta_0$ is the bulk
condensate, and $\psi$ is the electron field in the graphene layer;
here only the anomalous terms have been retained.   Upon
diagonalizing the quadratic Hamiltonian for $\psi$, a graphene
condensate $\Delta \ \equiv \ \langle \psi \psi \rangle$ is induced.

That broad-brush sketch took no notice of the intricate internal
structure of the graphene field $\psi$ and (possibly) of $\Delta_0$
and consequently $\Delta$.   The relevant, low-energy modes of
$\psi$ are labeled by a 2-momentum $k$ and three binary indices. The
first binary index is the valley index, which specifies whether the
mode arises from expansion around total momentum $p \ = \ \pm K +
k$; here $\pm K$ are the two momenta where the bands touch.   The
second binary index labels a pseudospin, arising from the
non-trivial residual symmetry of the unit cell, which roughly
speaking specifies whether the electron is on the A or B sublattice
of the bipartite honeycomb lattice.  This pseudospin is very
significant dynamically, as it appears in the effective Dirac
equation for these modes (see below).   Finally, the third binary
index labels ordinary spin.

As in helium-3 \cite{helium3Review}, or for that matter QCD
\cite{qcdReview}, internal structure for the fermion field opens up
many possibilities for the form of condensation.  For definiteness,
let us assume that the bulk condensate pairs electrons of opposite
{\it total\/} 2-momentum.  (This excludes LOFF-type bulk
superconductivity \cite{LOFF}.)  Then only intervalley pairing of
the form
\begin{equation}
\Delta_{+**;-**} (k ) \ \equiv \langle \psi_{+**} (k) \psi_{-**} (-k) \rangle
\end{equation}
is induced; intravalley pairing requires $\Delta_0 (\pm 2K ) \ \neq
0$.  (The possibility of intrinsic intravalley pairing has been
discussed \cite{intravalley}.)   Here the three indices are the
respective binary indices mentioned previously and $*$ is a
wildcard.   A general restriction arises from Fermi statistics, but
it is very weak: components of the condensate which are overall
antisymmetric under interchange of spatial, pseudospin, and spin
must be symmetric in the valley index, and {\it vice versa}.

With different bulk superconductors, one can imagine many exotic
possibilities being realized, e.g. gapless or
gapped $p$-wave or $d$-wave pairing
(or LOFF states \cite{LOFF}).    The most conventional
superconductors, however, are T-invariant, $s$-wave, and spin singlet.
Assuming that the primary interaction in Eqn. (\ref{interaction})
conserves spin (as is appropriate at least for light elements), then
the induced superconductivity will likewise be $s$-wave spin
singlet, and therefore symmetric in space and antisymmetric in spin.
This leaves two possibilities: symmetric in both valley and
pseudospin; or antisymmetric in both valley and pseudospin.   The
first possibility encompasses $3\times 3 = 9$ components, the second
just 1.    Finally, although there is not full rotational invariance
in pseudospin, the underlying $C_{6v}$ symmetry of the honeycomb
lattice on a homogeneous substrate is enough to insure that the induced, invariant quadratic
term is antisymmetric, i.e. pseudospin singlet in the usual sense.  {}For the decomposition of the tensor product of a 2-dimensional spinor representation of $C_{6v}$ with itself contains the identity representation of that group only once.

The preceding discussion pointed to many byways worthy of further
investigation. For our present purpose, however, the central
conclusion is that conventional bulk superconductors will induce a
very specific form of condensate, antisymmetric in each of the
internal indices, describable by a single complex-number field.   We
shall adopt that choice, implicit in \cite{beenakker}, in what
follows.


\section{Multivortices and Near-zero Modes}


The BdG equations from \cite{beenakker}, extended to include an
electromagnetic vector potential, take the form
\begin{equation}\label{bdgNoZeeman}
\left(\begin{array}{cccc} H^p_++H^A_+ -E &0 & \Delta(\textbf{r})& 0 \\
0 & H^p_-+H^A_-  - E &  0 & \Delta(\textbf{r})\\ \Delta^*(\textbf{r}) & 0
& -H^p_++H^A_+ - E &  0 \\ 0 & \Delta^*(\textbf{r}) & 0 & -H^p_-+H^A_- -  E
\end{array} \right) \left(\begin{array}{ccc}
u_+ \\ u_- \\v_-\\v_+ \end{array} \right)=  0
\end{equation}
where we absorb the coupling constant into $\Delta$, put
$\hbar=v_f=1$, measure energies relative to the cone apices, and
write $H_\pm = H^p_\pm+H^A_\pm $,  defining $H^p_\pm \equiv
-i(\sigma_x
\partial_x \pm \sigma_y \partial_y)$ and $H^A_\pm \equiv -q(\sigma_x A_x\pm \sigma_y A_y)$.  Here the subscripts refer to the valley index and the internal indices, the internal indices are for pseudospin, $u,v$ refer to particle and hole modes, respectively.    In this approximation ordinary spin is taken to be dynamically inert, and does not appear.   Note that the condensate mixes spin up particles with spin down holes, and {\it vice versa}. 

These equations decouple into two independent sets, one involving
$(u_+, v_-)$ and $H_+$, the other those objects with the
complementary indices. The equations for the second set can be
related to those for the first by reflection about the $x$ axis.
Restricting to the first set, dropping the indices, and focusing on
$E=0$, we find the equations
\begin{eqnarray}\label{godzilla}[-i(\sigma_x
\partial_x + \sigma_y \partial_y)-q(\sigma_x A_x+ \sigma_y A_y)]\ u+\Delta(\textbf{r})\ v&=0 \\
\label{rodan}\Delta^*(\textbf{r})\ u+[i(\sigma_x
\partial_x + \sigma_y \partial_y)-q(\sigma_x A_x+ \sigma_y A_y)]\ v&=0
\end{eqnarray}
Now putting $v = \sigma_y u^*$ we find that the second equation
reduces to the complex conjugate of the first, which reads
\begin{equation}\label{twoComponentDirac}
[-i(\sigma_x \partial_x + \sigma_y \partial_y)-q(\sigma_x A_x+ \sigma_y A_y)]\ u+\Delta(\textbf{r})\ \sigma_y u^* =0
\end{equation}
Equation (\ref{twoComponentDirac}) is precisely the equation for
zero modes of the relativistic two-component Dirac equation studied
in Refs. \cite{jr, eWeinberg}.

For later use, and to make our discussion self-contained, let us
briefly review the solutions of Eqn. (\ref{twoComponentDirac}) for
multivortices.   With $\textbf{A}=-\hat{e}_\theta A(r)$, and using
polar coordinates, we find for the upper component of $u$ -- call it
$a$ -- the equation
\begin{equation}\label{aEquation}
e^{i\theta} \bigl( \frac{\partial}{\partial r} + \frac{i}{r} \frac{\partial}{\partial \theta} \bigr) a - q A e^{i\theta} a +  \Delta a^* \ = \ 0
\end{equation}
and for the lower component, $b$, the equation
\begin{equation}\label{bEquation}
-e^{-i\theta} \bigl( \frac{\partial}{\partial r} - \frac{i}{r} \frac{\partial}{\partial \theta}  \bigr) b + q A e^{-i\theta} b -  \Delta b^* \ = \ 0
\end{equation}
Let us focus on the former.   First, we can remove the vector
potential term by rescaling $a = \exp ( q \int^r_0 A(s) ds ) \tilde
a$:  For a vortex $A$ is vanishes at the origin, and $A$  also vanishes
$\propto r^{-1}$ as $r\rightarrow \infty$, so the redefinition will not
affect the normalizability of the zero-modes we find below, which die
exponentially at infinity.  

We suppose that $\Delta (r, \theta) = \Delta_{(n)} (r)
e^{in\theta}$, with $\Delta_{(n)} (r) \rightarrow r^{|n|}$ as $r
\rightarrow 0$ and $\Delta_{(n)} (r)\rightarrow {\rm const.} $ as
$r\rightarrow \infty$, as is appropriate to an $n$-fold multivortex.  (In this situation $A(r) \rightarrow -\frac{n}{2q}r^{-1}$, so the vector potential asymptotically ``pulls in'' $\tilde a$ by a power of half the vorticity.)   
The definite partial wave solutions of Eqn. (\ref{aEquation}) are of
two types.   One type involves a single angular dependence, $\tilde
a = f(r) e^{il\theta}$.    The consistency condition for $l$ is
easily found to be $2l = n-1$.  The second type involves two angular
dependencies, $\tilde a = f(r) e^{il\theta} + g(r) e^{im\theta}$.
The consistency condition for $l$ and $m$ is $l+m = n - 1$.

For the first type, we derive the radial equation
\begin{equation}
f^\prime - \frac{l}{r} f + \Delta_{(n)} f^* \ = \ 0
\end{equation}
Assuming for simplicity that $\Delta_{(n)}$ is real, and positive at
infinity, $f$ can be taken real.  At large $r$ the middle term can
be dropped, and the solution dies exponentially.  At small $r$ the
last term can be dropped, and we see that solution is normalizable
for $l \geq -\frac{1}{2}$, or, since $l$ is integral, $l \geq 0$.
For the second type, we derive the radial equations
\begin{eqnarray}
f^\prime - \frac{l}{r} f + \Delta_{(n)} g^* \ &= \ 0 \\
g^\prime - \frac{m}{r} g + \Delta_{(n)} f^* \ &= \ 0
\end{eqnarray}
Now there are both growing and dying modes at infinity.   In order
to assure that the dying mode matches onto a normalizable solution
at the origin, we must require both $l, m \geq 0$.  With the same
assumptions on $\Delta_{(n)}$, there is a solution $f_0, g_0$ with
$f$ and $g$ both real and in phase at infinity, and another $if_0,
-i g_0$ with $f$ and $g$ pure imaginary and out of phase.   On the
other hand, from the definition interchange of $l$ and $m$ is a
trivial operation. Thus for positive $n$ we find an $n$-real
dimensional manifold of zero modes.   When $n$ is odd there is one
solution of the first type, corresponding to $l = \frac{n-1}{2}$,
and $n-1$ of the second type, corresponding to $n-1 \geq l \geq 1$.
When $n$ is even all solutions are of the second type, with $n-1
\geq l \geq 0$.

A very similar analysis applies to the $b$ equation, Eqn.
(\ref{bEquation}). In that case, there are $|n|$ normalizable modes
for $n \geq 0$, and none for $n \leq 0$. In all cases, the number of
solutions of the $a$ equations minus the number of solutions of the
$b$ equations equals $n$, the vorticity.    This suggests the
existence of an underlying index theorem, since index theorems
generally express the difference between the number of zero modes of
a differential operator $\cal D$ and the number of solutions of its
adjoint $ \cal D^\dagger$ -- by definition, the index of $\cal D$ -- in terms of
topological data in the structure of $\cal D$ \cite{indexReview}.
Index theorems can be very valuable in physics, because they reveal
the presence of low-energy modes whose existence might otherwise be
hard to anticipate, and insure that the existence of such low-energy
modes is robust against many kinds of perturbations (or
approximations).   Following E. Weinberg, if we write for the upper
component $a = h + ik$, with $h$ and $k$ real, and similarly $\Delta
= \phi_1 + i \phi_2$, and combine $h,k$ into a two-component spinor,
then the Dirac equation Eqn. (\ref{twoComponentDirac}) takes the
form
\begin{equation}
0 \ = \ \Bigl( (\frac{\partial}{\partial y} - q A_x ) + \tau_1 \phi_2 + i\tau_2 (\frac{\partial}{\partial x} + q A_y )\Bigr) \left(\begin{array}{c} h \\ k \end{array}\right) \ \equiv \cal{D} \left(\begin{array}{c} h \\ k \end{array}\right)
\end{equation}
while for the lower component, similarly decomposed, we find the
adjoint equation.  There is indeed an index theorem for $\cal D$,
essentially equating the index to the vorticity \cite{eWeinberg}.

The equations involving $(u_-, v_+)$ and $H_-$ contribute another
$|n|$ zero modes, after a parallel analysis.   Finally there is
another overall doubling, when we restore the physical spin
variable.

So far our discussion has been based on the approximate BdG equation
(\ref{bdgNoZeeman}), which does not include the Zeeman coupling of
spin (as opposed to pseudospin) to the magnetic field.   This is a
small effect quantitatively, but it has significant qualitative and
conceptual implications.   The Zeeman coupling makes an additional
diagonal contribution $\pm \kappa B {\bf 1} $ to the matrix in
(\ref{bdgNoZeeman}), where $B$ is the magnetic field, with the sign
depending on spin.   (Intuitively: since the background Cooper pairs
are spin singlets, mixing with them does not affect the Zeeman
energy.)   As a result the former zero-energy states are shifted. In
first-order perturbation theory in $\kappa$, we have the shifts
\begin{equation}
\epsilon_\pm \ = \ \pm \kappa \frac{\int_0^\infty dr 2\pi r B(r) | a(r) |^2 }{\int_0^\infty dr 2\pi r | a(r) |^2 }
\end{equation}
Thus it is roughly proportional to the average magnetic field over
the mode.   For a crude estimate of the ``minimal'' splitting, take
the product of the bare $g$-factor of an electron and the quantized
minimal quantum fluxoid spread over a square micron, then $\epsilon
\sim  10^{-7}$ eV, i.e. 1 mK.   Larger splittings might be obtained
using in-plane magnetic fields, which couple to spin but do not
frustrate the superconducting order parameter.

It might seem paradoxical that a small perturbation can shift ---
and thus remove, as strict zero modes --  zero modes whose existence
was tied to topology.   In the present context, however, inclusion
of the Zeeman term blocks passage from the amended Eqns.
(\ref{godzilla}, \ref{rodan}) to Eqn. (\ref{twoComponentDirac}), for
which the index theorem applies.    Alternatively, we could formally
include a Zeeman coupling-like term directly in Eqn.
(\ref{twoComponentDirac}).   We would then still find zero modes to
satisfy the index theorem for the slightly perturbed $\cal D$, but
we could not use them to get zero modes of the amended Eqns.
(\ref{godzilla}, \ref{rodan}).

\bigskip

\section{Comments}

\bigskip

\begin{enumerate}
\item It is convenient to have a single word to convey the concept ``soliton that acquires internal structure due to existence of localized low-energy modes of quantum fields it interacts with''.   We propose the term {\it modicule}, pronounced mode-icule, in view of the resemblance of such entities to emergent molecules.
\item {\it Experimental probes}:  General techniques for probing the internal structure of vortices and multivortices, notably including tunneling microscopy, were outlined in the pioneering work of Virtanen and Salomaa \cite{VS}.   As they emphasized, multivortices can be encouraged to form at defects.   The characteristic Zeeman splittings, predicted above, might be probed in absorption.   (Note that the spatial wave functions for opposite spins are accurately matched.)
\item {\it Ragged multivortices and edges}: Because the existence and number of zero modes is governed by an index theorem, and the relevant topological information is insensitive to the existence even of large holes where $\Delta$ vanishes, we will have near-zero modes associated with the edge of an annular region of graphene superconductor threaded by magnetic flux.   Nor is symmetry required.
\item {\it Quantum statistics}: For odd $n$, and in particular for the unit vortex $n=\pm 1$, we have unpaired, essentially real (``Majorana'')  modes.    There are four of them, due to the spin and valley degeneracies.   To a first approximation they are dynamically independent, because the spin couples feebly and intervalley scattering requires large momentum exchange $\pm 2K$.   If we quantize them separately we would find following Ivanov \cite{nonabelian} four copies of the Clifford algebra discovered by Nayak and Wilczek \cite{nw}.    As a consequence the exchange operation, which acts as 
\begin{eqnarray}
\gamma_j ~&\rightarrow&~ \gamma_k \nonumber \\
\gamma_k ~&\rightarrow&~ - \gamma_j
\end{eqnarray}
for a single mode, reduces to the trivial (non-entangling)
\begin{eqnarray}
\gamma_j^1 \otimes \gamma_j^2 ~&\rightarrow&~ \gamma_k^1 \otimes \gamma_k^2 \nonumber \\
\gamma_k^1 \otimes \gamma_k^2 ~&\rightarrow&~ \gamma_j^1 \otimes \gamma_j^2
\end{eqnarray}
for two, or any even number.  
\item {\it Comparing p + ip}: An effective Dirac equation of similar structure appears in the theory of superconductors with bulk $p_x + ip_y$ pairing \cite{pip}.    The approximation underlying it, however, is quite different.   Specifically, the momentum dependence is assumed to arise from the gap parameter, by expanding locally $\Delta (r , p) \rightarrow | \Delta(r)| ( e^{i\theta (r) } (p_x +i p_y) ) $ in the Bogoliubov-deGennes equation, where $\theta$ is the phase of the order parameter.   There are two potential problems with that approximation: it applies only if the ordinary momentum dependence can be neglected, and it breaks down when $|\Delta (r)|$ vanishes.  
\end{enumerate}

\end{document}